\documentclass[twocolumn,preprintnumbers,amsmath,amssymb,prl]{revtex4}
\usepackage{graphicx}
\usepackage{dcolumn}
\usepackage{bm}
\usepackage{units}%
\usepackage{stmaryrd}%
\usepackage[breaklinks=true,pdfborder={0 0 0}]{hyperref}
\usepackage[usenames]{color}
\usepackage{tabularx}

\usepackage{color}
\usepackage{tikz}
\usepackage{multirow}
\hypersetup{colorlinks=true,citecolor=red}

\newlength{\LL} \LL 1\linewidth
\begin{document}
\title{Fine-Structure Constant Connects the Polarizability of Atoms and Vacuum}
\author{Alexandre Tkatchenko}
\email{alexandre.tkatchenko@uni.lu}
\author{Dmitry V. Fedorov}
\affiliation{Department of Physics and Materials Science, University of Luxembourg, L-1511 Luxembourg City, Luxembourg}
\date{\today}
\begin{abstract}
We examine the recently derived quantum-mechanical relation between atomic polarizabilities and
equilibrium internuclear distances in van der Waals (vdW) bonded diatomic systems
[Phys.~Rev.~Lett. {\bf 121}, 183401 (2018)]. For homonuclear dimers, this relation is described by
the compact formula $\alpha_{\rm m}^{\rm q} = \Phi R_{\rm vdW}^7$, where the constant factor in front
of the vdW radius was determined empirically. Here, we derive
$\Phi = (4\pi\epsilon_0/a_0^4) \times \alpha^{\nicefrac{4}{3}}$ expressed in terms of the vacuum
electric permittivity $\epsilon_0$, the Bohr radius $a_0$, and the fine-structure constant $\alpha$.
The validity of the obtained formula is confirmed by estimating the value of the fine-structure
constant from non-relativistic quantum-mechanical calculations of
atomic polarizabilities and equilibrium internuclear vdW distances. The presented derivation
allows to interpret the fine-structure constant as the ratio between the polarizability densities of
vacuum and matter, whereas the vdW radius becomes a geometrical length scale of atoms
endowed by the vacuum field.
\end{abstract}
\maketitle

Electric polarization, which arises in atoms, molecules, and materials via formation of
permanent or transient electric multipoles, is a fundamental
property of matter. Similarly, quantum fields possess transient particle/antiparticle pair
fluctuations or \emph{virtual excitations},
which can be characterized by finite polarizability density~\cite{Rafelski1985,Milonni_book}.
For instance, the electric permittivity of vacuum is
$\epsilon_0 \approx 8.85 \times 10^{-12}$~F/m~\cite{Comment_1}. With units of polarizability
divided by volume~\cite{Atkins&Friedman_book}, $\epsilon_0$ is a key quantity to determine
the interaction between charges and polarization processes in matter in both classical
physics and quantum mechanics.

Within the classical picture, an atom, as a unit of matter (m), is described by a hard sphere
with its volume
$V_{\rm m} = 4\pi R_{\rm m}^3/3$ filled by
a homogeneous electron density surrounding an infinitely-heavy positive point charge.
The polarizability of such an atom can be obtained from the dipole moment induced
by an external electric field~\cite{Atkins&Friedman_book,Fedorov2018},
\begin{align}
\alpha_{\rm m}^{\rm cl}(R_{\rm m}) = (4\pi\epsilon_0) R_{\rm m}^3\ ,
\label{eqAlphaCM}
\end{align}
which implies that the polarizability density of matter is given by
$\alpha_{\rm m}^{\rm cl}/V_{\rm m} = 3 \epsilon_0$.
Thus, in a classical picture the polarizability of an atom is intimately connected to 
its Euclidean volume. However, for real atoms composed of many electrons interacting with a positively-charged 
nucleus and possessing an inhomogeneous electron density, the atomic radius $R_{\rm m}$ is not observable.

A proper quantum-mechanical description implies that a radius of an isolated atom
is strictly infinite, since its electron density (although decaying exponentially) is not confined 
in Euclidean space. Nevertheless, the polarizability is known to increase with an increasing number 
of electrons for a fixed nuclear charge.
Therefore, it is natural to search for an observable atomic size that
can be used to predict the polarizability of an atom. In our recent work~\cite{Fedorov2018}, 
we accomplished this task by obtaining a compact quantum-mechanical relation between the 
atomic polarizability and the vdW radius (half of the equilibrium internuclear distance in homonuclear dimers),
\begin{align}
\alpha_{\rm m}^{\rm q}(R_{\rm vdW}) = \Phi R_{\rm vdW}^7\ .
\label{eqAlphaQM}
\end{align}
The prefactor $\Phi$ was found to be a \emph{universal} constant with the value of $(2.54 \pm 0.02)^{-7}$ in atomic units (a.u.), as
empirically determined by the analysis of reference polarizabilities and vdW radii for the noble gases~\cite{Fedorov2018}.
We have shown that essentially the same constant remains valid for 72 atoms in the periodic table for 
which the vdW radii are known from experimental crystal structure data. This number of atoms covers 
about 60\% of all known chemical elements and corresponds to the intervals
of the atomic polarizability and the vdW radius spanning $1.38 \leq \alpha_{\rm m} \leq 427.12$ and
$2.65 \leq R_{\rm vdW} \leq 6.24$ in a.u.~\cite{Fedorov2018}, respectively.

The scaling law $\alpha_{\rm m}^{\rm q} \propto R_{\rm vdW}^7$ of Eq.~(\ref{eqAlphaQM}) is robust.
It was derived~\cite{Fedorov2018} from the balance between
quantum exchange and correlation forces in closed-shell diatomic systems described within
the quantum Drude oscillator model~\cite{Jones2013}. In addition, the same
power law was obtained~\cite{Fedorov2018} from the Tang-Toennies model of
the interatomic potential for the vdW-bonded dimers~\cite{Tang1984}.
The quantum expression for the polarizability scales as $R_{\rm vdW}^7$
in stark contradiction to the classical power law $R_{\rm m}^3$
of Eq.~(\ref{eqAlphaCM}). The large difference in the scaling laws is
incidentally compensated by the respective coefficients. In atomic units,
\begin{align}
\Phi \approx (4 \pi \epsilon_0 / a_0^4) \times (2.54)^{-7}
\label{eqPhi_2.54}
\end{align}
is approximately equal to $0.001466$ being much smaller than $4\pi\epsilon_0=1$.
To clarify the origin of the coefficient $\Phi$ and the scaling law $R_{\rm vdW}^7$
is the main aim of this work.

In this Letter, we derive $\Phi$ by establishing a relation between
the polarizability density of the vacuum field and that of matter.
The constant of Eq.~(\ref{eqPhi_2.54}) turns out to be related to the fine-structure constant (FSC) $\alpha$.
This is remarkable because in non-relativistic quantum mechanics, described by
the Schr\"odinger equation, the FSC tends to zero because the 
speed of light $c$ tends to infinity~\cite{Strange_book}. Hence, the FSC enters formulas
in a product with the speed of light as $\alpha c$, which is trivially equal to unity in
atomic units. To establish the validity of our derivation, we use atomic polarizabilities
and equilibrium distances in homonuclear dimers of noble gases computed from non-relativistic
quantum-mechanical treatment of valence electrons. Overall, our results suggest an intimate
relation between polarizabilities of matter and the underlying vacuum field, yielding
insights into the atomic quantum geometry.

We start by expressing the well-known formula for the FSC as a ratio between two polarizability densities
\begin{align}
\label{eqFSC}
\alpha = \frac{e^2 / \hbar c}{(4\pi\epsilon_0)} = \frac{\alpha_1 / V_1}{\alpha_2 / V_2}\ ,
\end{align}
where $\alpha_{1(2)}$ and $V_{1(2)}$ are the polarizability and volume of two different elements.
Given that the FSC measures the interaction between (real or virtual) photons and charged matter particles,
we associate one of the polarizability terms to the vacuum field and another one to matter.
Let us fix a volume of the field element $V_{\rm f} = V_1$ and obtain the associated
field polarizability $\alpha_{\rm f} = \alpha_1$. Then, determine a volume of the matter element 
$V_{\rm m} = V_2$, such that $\alpha_{\rm m} = \alpha_{\rm f}$. 
Under this condition, Eq.~(\ref{eqFSC}) can be rewritten as
\begin{align}
\label{eqFSCV}
\alpha = \left( V_{\rm m} / V_{\rm f} \right)\big|_{\alpha_{\rm m} = \alpha_{\rm f}}\ .
\end{align}
The placing of $V_{\rm m}$ in the numerator and $V_{\rm f}$ in the denominator is unambiguous and chosen such that
the polarizability density of matter becomes larger than that of the vacuum field for a fixed volume by
$\alpha^{-1} \approx 137.036$. Equation~(\ref{eqFSCV}) can also be written in terms of the radii of
the field and matter volume elements
\begin{align}
R_{\rm f}^3 = \alpha^{-1} R_{\rm m}^3\ \ \ \text{and}\ \ \ R_{\rm f} = \alpha^{-\nicefrac{1}{3}} R_{\rm m} \quad ,
\label{R_FandR_M}
\end{align}
under the condition of equal polarizabilities. The physical interpretation of these equations is the following.
The polarizability of a vacuum sphere of radius $R_{\rm f}$ is $\alpha^{-1}$ times smaller than
the polarizability of a sphere of matter with the same radius. Our forthcoming analysis confirms that $\alpha$
describes the ratio of polarizability densities of vacuum and matter, enriching the existing spectrum of textbook
interpretations of the FSC~\cite{Strange_book}.

As mentioned above, the quantum-mechanical polarizability of matter $\alpha_{\rm m}^{\rm q}$ 
given by Eq.~(\ref{eqAlphaQM})
has been derived based on the balance between forces stemming from 
quantum exchange and correlation interactions~\cite{Fedorov2018}.
The exchange force arises from the spin-statistics theorem in quantum-field theory~\cite{Dirac_book,Pauli1950},
which provides further evidence that $\alpha_{\rm m}^{\rm q}$ is determined by the interaction
between matter and the vacuum field. The classical and quantum expressions for the polarizability
of matter, Eqs.~(\ref{eqAlphaCM})--(\ref{eqAlphaQM}), intersect at one and only one
critical radius $R_{\rm c}$ determined by the vacuum field.
According to Eq.~(\ref{R_FandR_M}), we define $R_{\rm c} = \alpha^{-\nicefrac{1}{3}} a_0$, where $a_0$
is the characteristic atomic length scale. Imposing the condition
\begin{align}
\alpha_{\rm m}^{\rm cl}(R_{\rm c}) = \alpha_{\rm m}^{\rm q}(R_{\rm c})\ ,
\label{eqBalance}
\end{align}
we obtain the polarizability of matter as $(4\pi\epsilon_0) \alpha^{-1} a_0^3$.
According to Eqs.~(\ref{eqFSC}) and (\ref{eqFSCV}), the corresponding field polarizability
turns out to be unitary (in atomic units).
Finally, using Eq.~(\ref{eqBalance}), we identify the proportionality constant in
Eq.~(\ref{eqAlphaQM}) as
\begin{align}
\Phi = (4 \pi \epsilon_0 / a_0^4) \times \alpha^{\nicefrac{4}{3}}\ .
\label{eqPhi}
\end{align}
Thus, the numerical coefficient of $(2.54)^{-7}$ in Eq.~(\ref{eqPhi_2.54}), which was empirically
fixed in Ref.~\cite{Fedorov2018} to its average value for the noble gas atoms
spanning the range $(2.54 \pm 0.02)^{-7} \approx 0.001388-0.001549$, is now specified
as $\alpha^{\nicefrac{4}{3}} \approx 0.001415$.

\begin{figure*}[t]
\includegraphics[width=3in]{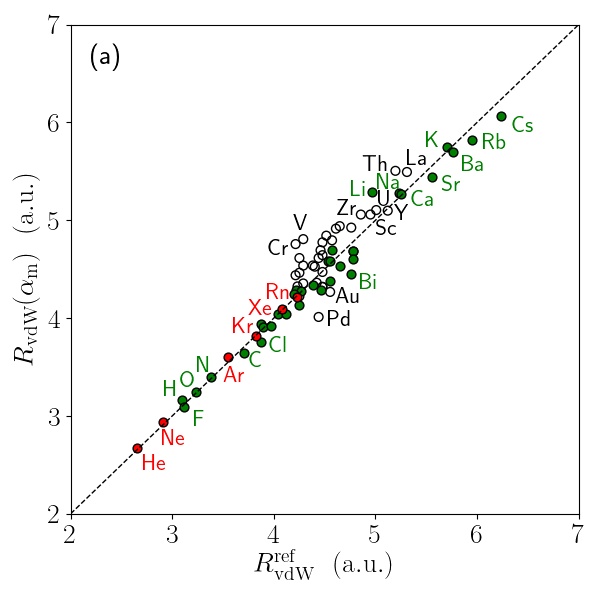} \qquad
\includegraphics[width=3in]{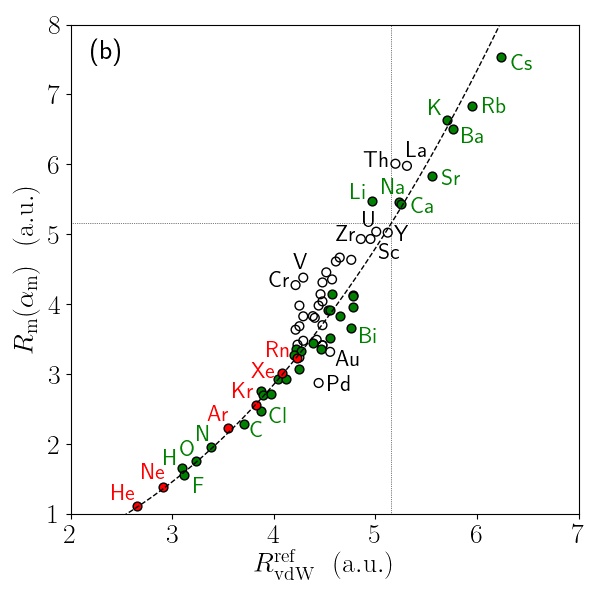}
\caption{The comparison between literature values for
$R_{\rm vdW}^{\rm ref}$ from the data of Refs.~\cite{Batsanov2001,Bondi1964} and
(a) atomic vdW radii $R_{\rm vdW} (\alpha_{\rm m})$ obtained by means of Eq.~(\ref{eqAlphaQMfinal})
or (b) the polarizability radii 
$R_{\rm m} (\alpha_{\rm m})$ calculated from Eq.~(\ref{eqAlphaCM}) for 72 atoms in the periodic table.
In the latter case, the dotted line corresponds to using
$R_{\rm vdW} (\alpha_{\rm m})$ instead of $R_{\rm vdW}^{\rm ref}$.
For these calculations, atomic polarizabilities of Ref.~\cite{Gobre2016} were
used and all quantities are presented in atomic units. The noble gases and the transition metals are indicated
by red and empty dots, respectively. All other elements including non-metals and simple metals are shown with
green dots.}
\label{figPlot}
\end{figure*}

The formula given by Eq.~(\ref{eqPhi}) is the central result of our work. Combining Eqs.~(\ref{eqAlphaQM}) and (\ref{eqPhi})
yields the final quantum-mechanical expression for the atomic polarizability in terms of three fundamental physical constants and the vdW radius
\begin{align}
\alpha_{\rm m}^{\rm q} = (4\pi\epsilon_0 / a_0^4) \alpha^{\nicefrac{4}{3}} R_{\rm vdW}^7\ .
\label{eqAlphaQMfinal}
\end{align}
Remarkably, the power $\nicefrac{4}{3}$ in the above equation can also be inferred from the relation between
the polarizability and volume of a quantum harmonic oscillator with 
arbitrary mass and frequency, specified as
$\alpha_{_{\rm QDO}} \propto V_{_{\rm QDO}}^{\nicefrac{4}{3}}$~\cite{Gobre2016,Kleshchonok2018}.
Indeed, $\alpha^{-\nicefrac{4}{3}} a_0^4$ can be represented as $(\alpha^{-1} a_0^3)^{\nicefrac{4}{3}}$,
where $\alpha^{-1} a_0^3$ has units of volume. In our recent work~\cite{Fedorov2018}, the coefficient $\Phi$
was empirically fitted to polarizabilities and interatomic distances of noble gas dimers. Now we can
reconsider this procedure for $\Phi$ derived in terms of the FSC
and given by Eq.~(\ref{eqPhi}). The statistical analysis in Ref.~\cite{Fedorov2018} was done
on the prefactor
$\Phi^{-\nicefrac{1}{7}}$ of the expression for $R_{\rm vdW}$ in terms of $\alpha_{\rm m}^{\rm q}$.
Hence, here we compare the corresponding $1/\alpha^{\nicefrac{4}{21}} \approx 2.5528$ derived from
Eq.~(\ref{eqPhi}) to the range $2.54\pm 0.02$ obtained in Ref.~\cite{Fedorov2018} for noble gas atoms.
Our analytical derivation falls well within the statistical error bar of the empirical determination with
a deviation of just $0.5\%$ from the mean.

A graphical illustration of the accuracy of Eq.~(\ref{eqPhi}) is provided by Fig.~\ref{figPlot}(a),
where $R_{\rm vdW} (\alpha_{\rm m})$ calculated by Eq.~(\ref{eqAlphaQMfinal}) from the accurate values 
for atomic polarizabilities~\cite{Gobre2016} are shown versus the experimental reference data from 
Refs.~\cite{Batsanov2001,Bondi1964}.
The atomic polarizabilities and the vdW radii are known for the noble gases with a very
high accuracy. This explains the excellent agreement between $R_{\rm vdW} (\alpha_{\rm m})$
and $R_{\rm vdW}^{\rm ref}$ in Fig.~\ref{figPlot}(a) for these atoms. A similar situation is
found for organic elements. For other atoms, the reference data is less accurate, because of
the increased statistical errors in the tabulated $R_{\rm vdW}^{\rm ref}$~\cite{Batsanov2001}.
The worst case corresponds to the transition metals possessing various spin states in the crystal
compounds used for the evaluation of their vdW radii from the experimental data. This is the reason for
the largest deviations between $R_{\rm vdW} (\alpha_{\rm m})$ and $R_{\rm vdW}^{\rm ref}$ obtained for such
chemical elements as Cr (12.88\%) and V (12.07\%). However, the average relative error for all 72 elements
shown in Fig.~\ref{figPlot}(a) is just 3.07\% and it reduces to 0.05\% by restricting our
consideration to the noble gases only. Thus, the main reason for the observed discrepancies is caused by
the statistical error in determination of $R_{\rm vdW}^{\rm ref}$~\cite{Batsanov2001}. In addition,
there might be a small influence from higher-order quantum electrodynamic contributions to the atomic
polarizabilities, scaling as higher powers of the FSC. Altogether, this analysis validates our
derivation of the intimate connection between matter and vacuum polarization processes.

In addition to the discussion of the two vdW radii shown in Fig.~\ref{figPlot}(a), we compare
the radii $R_{\rm m} (\alpha_{\rm m})$ calculated by means of Eq.~(\ref{eqAlphaCM}) to $R_{\rm vdW}^{\rm ref}$
from the experimental reference data collected in Refs.~\cite{Batsanov2001,Bondi1964}.
This comparison is shown in Fig.~\ref{figPlot}(b), where the dotted line corresponds to the use of
$R_{\rm vdW} (\alpha_{\rm m})$ obtained from Eq.~(\ref{eqAlphaQMfinal}) instead of $R_{\rm vdW}^{\rm ref}$,
which contain large statistical errors for elements such as transition metals~\cite{Batsanov2001}.
The value of the two radii, at which Eqs.~(\ref{eqAlphaCM}) and (\ref{eqAlphaQMfinal}) possess the same
polarizabilities and radii, 
$\alpha_{\rm m}^{\rm cl} = \alpha_{\rm m}^{\rm q} = (4\pi\epsilon_0) \alpha^{-1} a_0^3 
\approx 137.036$~a.u.~and $R_{\rm m} = R_{\rm vdW} = \alpha^{-\nicefrac{1}{3}} a_0 \approx 5.1556$~a.u., 
is indicated by the two intersecting lines in Fig.~\ref{figPlot}(b).
Similar to Fig.~\ref{figPlot}(a), most atoms in the periodic table are close to the dotted line,
confirming a simple non-linear relation between $R_{\rm m}$ and $R_{\rm vdW}$ involving the FSC.
As expected, noble gas atoms and organic elements, for which the statistical data for the determination
of the vdW radius are abundant, are particularly close to the dotted line. It is worth noting the
contrasting behavior of noble gases and alkali metal atoms in Fig.~\ref{figPlot}(b). While the noble
gas series He-Ne-Ar-Kr-Xe-Rn seems to slowly converge towards the ``magic'' radius 
of $R_{\rm c}=\alpha^{-\nicefrac{1}{3}} a_0$ from below, the alkali atoms Cs-Rb-K-Na-Li converge to 
this radius from above.
Remarkably, for many elements in the periodic table, $R_{\rm m}$ and $R_{\rm vdW}$ are
relatively close to $R_{\rm c}$.
This suggests that in the asymptotic case with respect to the atomic number
Eqs.~(\ref{eqAlphaCM}) and (\ref{eqAlphaQMfinal}) are seamlessly connected.
In other words, for macroscopic matter, the quantum-mechanical formula of Eq.~(\ref{eqAlphaQMfinal})
transforms to its classical counterpart given by Eq.~(\ref{eqAlphaCM}).

It is worth discussing the two radii considered above in more detail. The classical
atomic radius $R_{\rm m}$ defined by Eq.~(\ref{eqAlphaCM}) is also known as the polarizability
radius~\cite{Dmitrieva1983,DeKock2012}. This radius was introduced based on the fact~\cite{Hirschfelder1964}
that for a uniform conducting sphere of radius $R_{\rm m}$ the polarizability has the same expression
as given by Eq.~(\ref{eqAlphaCM}). The polarizability radius has been used in chemical literature together 
with several other radii~\cite{Politzer2002,DeKock2012} aimed to qualitatively 
represent atomic size and enable comparison between different atoms. 
The vdW radius was defined by Pauling~\cite{Pauling1960} and
Bondi~\cite{Bondi1964} as half of the distance between two (closed-shell) atoms of the same chemical element,
at which Pauli exchange repulsion and London dispersion attraction forces exactly balance 
each other~\cite{Nunzi2020}.
Based on this definition, in Ref.~\cite{Fedorov2018} we derived the scaling law of
Eq.~(\ref{eqAlphaQMfinal}) and reasonably estimated the constant $\Phi$, which is finally specified here
by Eq.~(\ref{eqPhi}). Our derivation suggests that the vdW radius corresponds to a relation between
the length scales in vacuum and matter since balanced forces occur when atoms interact through
the vacuum. In other words, one can assume that the vdW radius describes the atomic size endowed by
the vacuum field. Therefore, the vdW radius should not strongly depend on the atomic environment. 
Indeed, the vdW radius for noble gas atoms (except for the lightest He atom) remains practically the 
same in gas, liquid, and solid states of matter~\cite{Batsanov2012}.  

\begin{table}[t]
\caption{
Estimation of the FSC from non-relativistic~\cite{Comment_2} quantum-mechanical calculations on noble gas atoms
and their homonuclear dimers.
We employ atomic polarizabilities
$\alpha_{\rm m}$ from Ref.~\cite{Gould2016}, calculated  equilibrium vdW radii
$R_{\rm vdW}^{\rm calc}$ from Refs.~\cite{Laschuk2003,Slavicek2003,Shee2015}, and experimental vdW
radii $R_{\rm vdW}^{\rm exp}$ from Ref.~\cite{Bondi1964}. All these values are given in atomic units.
The predicted values of the FSC, $\alpha(R_{\rm vdW}^{\rm calc})$ and $\alpha(R_{\rm vdW}^{\rm exp})$,
can be compared to its experimental value $\approx 0.007297$~\cite{Davis2017,Parker2018,Clade2019}.}
\begin{ruledtabular}
\begin{tabular}{lccccc}
\mbox{Species} & \mbox{
$\alpha_{\rm m}$} & \mbox{
$R_{\rm vdW}^{\rm calc}$} & \mbox{
$R_{\rm vdW}^{\rm exp}$} & \mbox{
$\alpha(R_{\rm vdW}^{\rm calc})$} & \mbox{
$\alpha(R_{\rm vdW}^{\rm exp})$} \\
\hline
He & 1.38  & 2.81 & 2.65 & 0.00561 & 0.00763 \\
Ne & 2.67  & 2.92 & 2.91 & 0.00753 & 0.00766 \\
Ar & 11.10 & 3.56 & 3.55 & 0.00774 & 0.00786 \\
Kr & 16.80 & 3.82 & 3.82 & 0.00730 & 0.00730 \\
Xe & 27.20 & 4.11 & 4.08 & 0.00713 & 0.00741 \\
Rn & 32.20 & 4.18 & 4.23 & 0.00741 & 0.00696 \\
\end{tabular}
\end{ruledtabular}
\label{tabAlpha}
\end{table}

An interesting application of Eq.~(\ref{eqAlphaQMfinal}) is the possibility to obtain the FSC,
which is the dimensionless quantity $\alpha = e^2/(4\pi\epsilon_0 \hbar c)$ arising in relativistic
quantum mechanics~\cite{Strange_book}, by means of non-relativisic quantum-mechanical calculations. 
One just needs to determine the atomic polarizability $\alpha_{\rm m}$ and the equilibrium interatomic
distance $R_{\rm eq} = 2 R_{\rm vdW}$ in homonuclear vdW-bonded dimers. Both quantities can be
computed very accurately from state-of-the-art quantum-chemical methods for closed-shell atoms.
The resulting prediction of the FSC is shown in Table~\ref{tabAlpha} for the six noble gases
present in Fig.~\ref{figPlot}. The atomic polarizabilities and equilibrium internuclear distances
have been obtained from high-level calculations based on self-interaction corrected
time-dependent density-functional theory for the polarizability~\cite{Gould2016} and CCSD(T)
for binding energy curves~\cite{Laschuk2003,Slavicek2003,Shee2015}.
The computed values $R_{\rm vdW}^{\rm calc}$ do not account for nuclear quantum vibrations,
particularly important for the He dimer. For this reason, in Table~\ref{tabAlpha} we also provide
comparative estimates of $\alpha$ obtained using the experimentally determined vdW radii.
Both estimates of the FSC are in good agreement, except for deviations observed for He and Rn.
In the former case, nuclear quantum effects contributing to the experimental $R_{\rm vdW}^{\rm exp}$
seem to be the source of the difference. For Rn, the discrepancy is presumably caused by the strong
relativistic effects influencing both, the vdW radius and the polarizability. Indeed,
with the polarizability of 33.54~a.u.~obtained for Rn in Ref.~\cite{Gobre2016} and
$R_{\rm vdW}^{\rm exp} = 4.23$ a.u., the estimated FSC would become 0.007176. Overall,
there is good agreement between $\alpha(R_{\rm vdW}^{\rm calc})$ or $\alpha(R_{\rm vdW}^{\rm exp})$
from Table~\ref{tabAlpha} and the well-known reference value $\alpha \approx 0.007297$ confirmed by
recent measurements~\cite{Davis2017,Parker2018,Clade2019}. The small
uncertainty in the values of the polarizabilities and the vdW radii is not the only reason for the
observed discrepancy. It is also likely that the quantum-mechanical formula given by Eq.~(\ref{eqAlphaQMfinal})
corresponds to a leading-order approximation. Higher-order terms with respect to the FSC need to be
considered for furnishing a more accurate formula.

In summary, we presented the derivation of the prefactor in the quantum-mechanical
relation between the atomic polarizability and vdW radius, $\alpha_{\rm m}^{\rm q} = \Phi R_{\rm vdW}^7$.
This constant factor was shown to be given by the simple expression
$\Phi = (4\pi\epsilon_0/a_0^4)\times\alpha^{\nicefrac{4}{3}}$ in terms of the three fundamental
physical constants: the vacuum electric permittivity $\epsilon_0$, the Bohr radius $a_0$, 
and the fine-structure constant $\alpha$. The validity of the derived expression was illustrated by 
the comparison between the calculated and reference vdW radii. It was shown that the vdW radius 
describes an atomic size endowed by the vacuum field, which has an intimate connection to the classical 
atomic radius, $R_{\rm m} = (\alpha_{\rm m}^{\rm cl}/4\pi\epsilon_0)^{\nicefrac 13}$, 
via the polarizability $\alpha_{\rm m}$.
The quantum and classical formulas for the atomic polarizability intersect at a value 
$\alpha_{\rm m} = \alpha^{-1}$ in atomic units, which corresponds to the unitary polarizability 
of the vacuum field. 
Altogether, this indicates that the fine-structure constant describes the ratio between the polarizability
densities of the vacuum field and matter.
\\ \\
We acknowledge financial support from the European Research Council (ERC Consolidator Grant \lq\lq BeStMo\rq\rq)
and the Luxembourg National Research Fund (FNR CORE project \lq\lq PINTA\rq\rq).


\begin{thebibliography}{99}
\bibitem{Comment_1} Throughout the paper, all the values of different quantities are given in atomic units
with $\hbar = m_e = e = 4\pi\epsilon_0 = 1$. However, for clarity, all physical constants are kept 
explicitly in the formulas.
\bibitem{Comment_2} In the calculations, relativistic effects were included approximately for the core
electrons of Xe and Rn. However, the valence electrons predominantly determining the polarizability
and the equilibrium distance were calculated using the non-relativistic Schr{\"o}dinger equation. 
The contribution of core electrons to the polarizability and the equilibrium distance are negligible 
for Xe and Rn.
\end{thebibliography}
\end{document}